\documentclass[
 reprint,
 amsmath,amssymb,
 aps,
 prl
]{revtex4-2}

\usepackage{graphicx}
\usepackage{dcolumn}
\usepackage{amsmath,amssymb,amsfonts}%
\usepackage{amsthm,bm,bbold}%
\usepackage{hyperref}
\hypersetup{
	colorlinks,%
	citecolor=blue,%
	filecolor=blue,%
	linkcolor=blue,%
	urlcolor=blue
}
\usepackage{lipsum}
\newcommand{\ket}[1]{\left\lvert #1 \right\rangle}
\newcommand{\bra}[1]{\left\langle #1 \right\rvert}
\newcommand{\trod}{\tilde\rho_{0\cdots0,1\cdots1}(t)}
\newcommand{\rod}{\rho_{0\cdots0,1\cdots1}(t)}

\newcommand{\ketbra}[2]{\ket{#1}\!\!\bra{#2}}

\let\perptmp\perp
\renewcommand{\perp}{{\! \mathsmaller{\perptmp}}}

\begin{document}

\title{Revealing correlated noise with single-qubit operations}

\author{Balázs Gulácsi}
 \email{balazs.gulacsi@uni-konstanz.de}
\author{Joris Kattemölle}
\author{Guido Burkard}%
\affiliation{Department of Physics, University of Konstanz, 78457 Konstanz, Germany}

\date{\today}

\begin{abstract}
Spatially correlated noise poses a significant challenge to fault-tolerant quantum computation by breaking the assumption of independent errors. 
Existing methods such as cycle benchmarking and quantum process tomography can characterize noise correlations but require substantial resources. 
We propose straightforward and efficient techniques to detect and quantify these correlations by leveraging collective phenomena arising from environmental correlations in a qubit register. In these techniques, single-qubit state preparations, single-qubit gates, and single-qubit measurements, combined with classical post-processing, suffice to uncover correlated relaxation and dephasing. 
Specifically, we use that correlated relaxation is connected to the superradiance effect which we show to be accessible by single-qubit measurements.
Analogously, the established parity oscillation protocol  can be refined to reveal correlated dephasing through characteristic changes in the oscillation line shape, without requiring the preparation of complex and entangled states.
\end{abstract}

\maketitle


\emph{Introduction---}
Realizing functional and scalable quantum devices remains a formidable challenge, mainly due to decoherence. In quantum computing specifically, decoherence rapidly introduces errors into qubits 
thereby undermining the integrity of computational processes. 
To make scalable quantum computing feasible, addressing decoherence is essential, with quantum error correction widely recognized as necessary to counter its effects.

Quantum error-correcting codes distribute quantum information across many physical qubits in a way that enables error-reversal procedures to correct decoherence effects~\cite{RevModPhys.87.307}.
To execute error reversal, one continuously gathers information about which errors occurred without directly measuring the quantum states, processes this data classically, and then applies corrective quantum operations on the quantum data. 
Since error detection and correction are also prone to noise, the existence of quantum error-correcting codes does not guarantee the ability to store or compute with quantum information for an arbitrarily long time.
Fortunately, fault-tolerance threshold theorems show that it is nevertheless possible to counter 
decoherence by adding redundancy---a manageable increase in qubits and computation time---as long as the error rate per physical qubit stays below some critical value, which is called the noise threshold~\cite{shor1996fault,aharonov1997fault, knill1998resilient,aharonov2008fault}.
These theorems have underpinned the feasibility of scalable quantum computing and have guided experimental efforts to reduce error rates below threshold values in physical qubits~\cite{barends2014superconducting,rong2015experimental,ballance2016high,noiri2022fast}.


Noise in quantum systems is often correlated both spatially and temporally, rather than occurring locally~\cite{GHZ,Boter2020,Wilen2021,McEwen2022,Yoneda2023,schloer2019}. 
These correlated errors introduce dependencies that standard error-correcting codes are typically not designed to handle. However, provided that the correlated noise remains short-ranged, fault tolerance is still achievable, albeit with additional overhead and with adjustments of the threshold values~\cite{DBLP:journals/qic/AliferisGP06,PhysRevLett.96.050504,PhysRevA.71.012336}. In practice, the presence of correlations necessitates a more nuanced approach to detection and correction, an area that remains far from understood. Ultimately, the effectiveness of error correction relies on an error model that accurately captures the dominant types of noise and errors. Given the significant impact of noise correlations, understanding their spatial and temporal characteristics is crucial to advance quantum computational platforms.

Many recent works have provided valuable insights into spatially correlated noise; however, these efforts are often restricted to a single type of noise or few-qubit systems~\cite{recentcorr1,recentcorr2,recentcorr3,recentcorr4,bosco1}. Moreover, other approaches rely on methods, such as cycle benchmarking~\cite{cyclebench} and process tomography~\cite{white2020,berg2023probabilistic}, that require significant overhead and access to a complete quantum hardware and software stack, resulting in potentially lengthy design feedback loops.

In this Letter, we present simple and efficient methods to detect the presence and strength of correlated noise, as well as its correlation length. In contrast to existing methods, our approach requires minimal technical resources, comparable to those needed for measuring standard qubit relaxation ($T_1$) and decoherence ($T_2$) times, and are inherently platform-agnostic. This is possible by leveraging collective phenomena associated with spatial correlations~\cite{spacecorr}; effects akin to \emph{superradiance}---where radiation from a group of emitters is enhanced by the correlations---signal correlated decay. 
Similarly, the related concept of \emph{superdecoherence}, where collective behavior amplifies the degradation of quantum coherences, and the potentially emerging decoherence-free subspaces~\cite{DFS}, where collective behavior reduces this degradation, can be used to detect correlated dephasing.


\emph{Formalism---}We assume the Hamiltonian $H=H_S(t)+H_B+H_{\textrm{SB}}$, with $H_S(t)=H_0+H_{\mathrm{ct}}(t)$ describing the quantum processor, $H_B$ the correlated environments, and  
\begin{equation}
    H_{\textrm{SB}}=\sum_{\alpha,j} A_{\alpha,j}\otimes B_{\alpha,j},\label{eq:int}
\end{equation}
the system-bath interactions. Here, $A_{\alpha,j}$ is a Hermitian one-qubit system operator \footnote{We consider no direct couplings between the qubits (only one-qubit terms in $H_S(t)$), therefore it is reasonable to assume that the environments only couple to one-qubit terms.} to which the environments connect through the Hermitian bath operators $B_{\alpha,j}$. Greek indexes denote the qubit locations, while Latin letters are used for the different noise effects, unless stated otherwise. Within the weak coupling theory~\cite{gulacsi1,bosco1} that we will utilize in the coming discussion, the only relevant information from the environments are the bath correlation functions
\begin{equation}
    \langle B_{\alpha,j}(t)B_{\beta,k}(0)\rangle=\delta_{jk}\int_{-\infty}^\infty\frac{\textrm d\omega}{2\pi}\ S_{\alpha\beta,j}(\omega)e^{-i\omega t}.\label{eq:corr}
\end{equation}
We emphasize that in the weak coupling limit, this form describes general correlated noise without any additional assumptions.
The appearance of $\delta_{jk}$, which signifies the mutually uncorrelated nature of different noise effects, arises as a consequence of the weak coupling assumption~\cite{Addgen}.
Spatial correlations emerge from the self-interactions of a particular bath $j$ which is described by its noise power spectrum $S_{\alpha\beta,j}(\omega)$. 

We transform to the interaction frame with respect to the bare time-independent part of the Hamiltonian of the quantum processor containing a register of $N$ qubits, $H_0=\sum_\alpha\omega_\alpha Z_\alpha/2$, where $\omega_\alpha$ is the respective qubit frequency and $Z_\alpha$ is the Pauli-Z matrix acting on qubit $\alpha$. This leads to the state of the register to be expressed as $\tilde\rho(t)=e^{iH_0t}\rho(t)e^{-iH_0t}$, with $\rho(t)$ being the reduced state in the laboratory frame. The dynamics of the register's state in the interaction frame is derived by tracing out the environments, using the time-convolutionless framework of open quantum systems~\cite{supp}, which yields the quantum master equation
\begin{equation}
    \frac{\textrm d}{\textrm dt}{\tilde \rho}(t)=-i[\tilde H_{\textrm{ct}}(t),\tilde\rho(t)]+\sum_j\tilde D_j[t,\tilde\rho(t)]\label{eq:TCL2}.
\end{equation}
The commutator containing $\tilde H_{\textrm{ct}}(t)$ describes the local coherent control of the qubits. The second term describes the detrimental effects of the environments through the explicitly time-dependent dissipators $\tilde D_j[t,\tilde\rho(t)]$ that go beyond the Lindblad form, incorporating both non-Markovian and spatially correlated noise processes. The different noise contributions are additive due to Eq.~\eqref{eq:corr}, which provides the opportunity to study the competition between the potential local and nonlocal effects. The dissipators depend on the one-qubit system operator $A_{\alpha,j}$, appearing in Eq.~\eqref{eq:int}, the coherent control applied to the qubits~\cite{gulacsi1}, and the noise power spectrum $S_{\alpha\beta,j}(\omega)$. 


\emph{Correlated relaxation---}Applying our formalism, we first turn to transverse coupling to the environment. Without loss of generality, we assume the system operator $A_{\alpha,\mathrm{relax}}=X_\alpha$, the Pauli-X matrix acting on qubit $\alpha$, in $H_{\mathrm{SB}}$. This coupling allows energy exchange between the processor and its environment, leading to qubit relaxation. In the absence of control terms, the dissipator for relaxation is
\begin{equation}
\begin{split}
    \tilde D&_{\textrm{relax}}[t,\tilde\rho(t)]=-i\left[H_{XY}(t),\tilde\rho(t)\right]\\
+&\sum_{\substack{\alpha,\beta \\ i,j}}\gamma_{\alpha\beta}^{(ij)}(t)\left(\Pi^{(i)}_\beta\tilde\rho(t)\Pi^{(j)}_\alpha-\frac{1}{2}\{\Pi^{(j)}_\alpha \Pi^{(i)}_\beta,\tilde\rho(t)\}\right).\label{eq:relax}
\end{split}
\end{equation}
The first term is the contribution due to the Hamiltonian $H_{XY}$ induced by the environmental effects, which describes bath-mediated symmetric and antisymmetric (Dzyaloshinskii--Moriya) exchange interactions (for details on $H_{XY}$, see the Supplemental Material~\cite{supp}). The second term describes the correlated decay processes, with $i,j\in\{1,2\}$ and  $\Pi_\alpha^{(1)}=\Pi_\alpha^\dagger=(X_\alpha -iY_\alpha)/2$ and $\Pi_\alpha^{(2)}=\Pi_\alpha=(X_\alpha +iY_\alpha)/2$ the local raising and lowering operators, respectively. The time-dependent rates are calculated via $\gamma_{\alpha\beta}^{(ij)}(t)=\int_{-\infty}^\infty\textrm d\omega/2\pi\ S_{\alpha\beta}(\omega)F^{(ij)}_{\alpha\beta}(\omega,t)$, where $F^{(ij)}_{\alpha\beta}$ denotes the filter function of the particular rate~\cite{supp}. We recognize $\gamma_{\alpha\beta}^{(12)}(t)\equiv \gamma_{\alpha\beta}^{\uparrow}(t)$ as the correlated absorption rates and $\gamma_{\alpha\beta}^{(21)}(t)\equiv \gamma_{\alpha\beta}^{\downarrow}(t)$ as the correlated emission rates. The remaining contributions are the nonsecular terms with the rates obeying the constraint $\gamma_{\alpha\beta}^{(11)}(t)=[\gamma_{\beta\alpha}^{(22)}(t)]^*$, which ensures that the master equation preserves the Hermiticity of the density matrix. These terms are typically neglected under the secular approximation because the rates $ \gamma_{\alpha\beta}^{(11)}(t)$ oscillate much more rapidly than the absorption and emission rates. However, as the number of qubits increases, the contributions of the nonsecular terms can accumulate, rendering the secular approximation invalid; therefore, we avoid using it here.

The correlated relaxation dissipator in Eq.~\eqref{eq:relax} is reminiscent of the one appearing in the master equation used to describe the collective spontaneous emission of an ensemble of excited atoms, that is, superradiance~\cite{PhysRevA.2.2038,PhysRevLett.95.243602,natdicke}. 
This motivates us to study the evolution of the system's total energy $W(t)=\sum_\alpha\omega_\alpha\langle Z_\alpha\rangle(t)/2$, as well as the total intensity of the radiation that leaves the system,
\begin{equation}
    I(t)=-\frac{\textrm d}{\textrm dt}W(t)=-\frac{1}{2}\sum_{\alpha,j}\omega_\alpha\textrm{Tr}\left\{Z_\alpha\tilde D_j[t,\tilde\rho(t)]\right\}.\label{eq:intensity1}
\end{equation}
Here, we used that $\langle Z_\alpha\rangle(t)=\textrm{Tr}[Z_\alpha\rho(t)]=\textrm{Tr}[Z_\alpha\tilde\rho(t)]$, so that the derivative and the intensity can be obtained using Eq.~\eqref{eq:TCL2}. The contributions to the intensity are additive with respect to the different decoherence effects, whether they are local or nonlocal. Using the relaxation dissipator in Eq.~\eqref{eq:relax}, it is straightforward to calculate the intensity,
\begin{equation}
I(t)=\sum_\alpha\omega_\alpha\left[\gamma^\downarrow_\alpha(t)\langle\Pi^\dagger_\alpha\Pi_\alpha\rangle-\gamma^\uparrow_\alpha(t)\langle\Pi_\alpha\Pi^\dagger_\alpha\rangle\right]+I_{\textrm{corr}}(t).\label{eq:intensity2}
\end{equation}
The first term represents the sum of independent emission and absorption by the qubits. The negative sign indicates energy entering the system. In this sum, the local emission rates $\gamma^\downarrow_\alpha$ and absorption rates $\gamma^\uparrow_\alpha$ are understood to include all local contributions, even those arising independently of the correlated environments (e.g., from additional local baths). The second term in Eq.~\eqref{eq:intensity2} appears due to the build-up of pair correlations between the qubits. This is an initial-state dependent quantity that leads to superradiance or subradiance in the context of collective spontaneous emission of atoms~\cite{PhysRevLett.95.243602}. 
In terms of the qubits, it reveals the presence of correlated incoherent noise.

\begin{figure}[t]
\includegraphics[width=\linewidth]{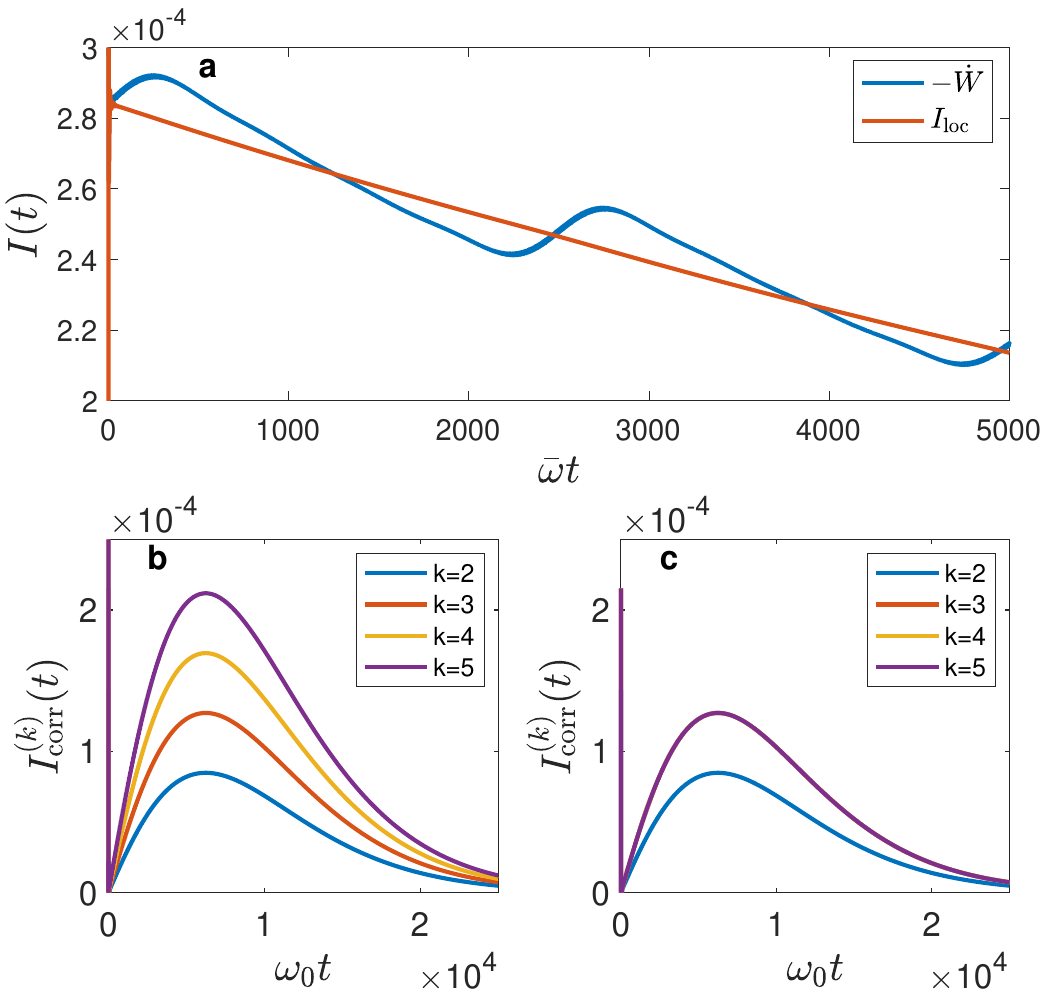}
\caption{\label{fig:1} The intensity of radiation leaving the $N=5$ qubit system in the presence of correlated relaxation. The strength of the Ohmic noise is chosen as $\lambda/\bar\omega=10^{-5}$ and the cutoff frequency as $\omega_c/\bar\omega=10$. (\textbf{a})
The derivative of the total energy (blue solid line) and the local contribution of the intensity (red solid line). Notice the blue line broadening due to the nonsecular terms. With the average frequency of the qubits $\bar\omega$ chosen as the unit of energy ($\hbar=1$), the qubit frequencies used on this plot are $\omega_\alpha/\bar\omega=0.9925+0.0025\alpha$ for $\alpha=1,\dots,5$. (\textbf{b}) The scaling of the correlated partial intensities with uniform qubit frequencies $\omega_0$ and bath correlation length encompassing all qubits. (\textbf{c}) As in (b), with the bath correlation length including only the first three qubits.}
\end{figure}
Measuring the intensity directly in quantum computing architectures would be impractical, as one typically does not have the means to place detectors into the processor's environment. Fortunately, this is not necessary. Due to the conservation of energy, by measuring the local expectation values $\langle Z_\alpha\rangle$ for each qubit over time, one can determine the total energy emitted from the system. These measurements are routinely performed, as the same data are commonly used in relaxation time ($T_{1\alpha}$) acquisitions. Remarkably, a more careful analysis of this data---specifically, postprocessing to obtain the derivative $\dot W(t)$---also reveals information about the spatial correlations of the noise.

A numerical example of the intensity is presented in Fig.~\ref{fig:1}. We consider $N=5$ qubits with no direct couplings between them and with average frequency $\bar\omega$, initialized in the fully inverted state $|11111\rangle$. We note that while uniform qubit frequencies amplify correlated effects, tuning the frequencies may not always be possible. For this reason, we first consider an example with unequal qubit frequencies to demonstrate that the correlated effects remain detectable even in this case (Fig.~\ref{fig:1}a).
The solution of the master equation in Eq.~\eqref{eq:TCL2} provides the register's state after idling time $t$ under the influence of correlated decay [Eq.~\eqref{eq:relax}]. The state $\tilde\rho(t)$ is used to calculate the time dependence of the total energy $W(t)$. For concreteness, here we assume a straightforward model for the spatial correlations: $S_{\alpha\beta}(\omega) = e^{i\theta} \xi_{\alpha\beta} S_{}(\omega)$, stemming from the general properties of the correlated noise spectrum \footnote{Generally, the local contribution of the noise spectrum is real-valued while the spatially correlated spectrum may be complex-valued, more importantly the correlated effects are always bounded by the local ones, $|S_{\alpha\beta,j}(\omega)|\leq S_{\alpha\alpha,j}(\omega)$.}. The phase $\theta$ and the real dimensionless factor $\xi_{\alpha\beta}\leq1$ determining the correlation length are generally functions of frequency, depending on the microscopic details of the environment. However, for simplicity, we assume that they are independent of frequency~\cite{bosco1}. 
We consider Ohmic relaxation at zero temperature, with an exponential cutoff and with $\xi_{\alpha\beta}=1$, so that $S_{\alpha\beta}(\omega\geq0)=\lambda\omega e^{-\omega/\omega_c}$ and $S_{\alpha\beta}(\omega<0)=0$, where $\lambda$ is the noise strength and $\omega_c\gg\bar\omega$ is the cutoff frequency. In the absence of correlations, only the independent emissions contribute to the intensity, which is well approximated by $I_{\textrm{loc}}(t)\approx\sum_\alpha\omega_\alpha\langle \Pi^\dagger_\alpha\Pi_\alpha\rangle/T_{1\alpha}$. Provided that the frequencies and the relaxation times of the qubits are known, the presence of correlated noise is signaled by  $I_{\mathrm{corr}}=-\dot W-I_{\mathrm{loc}}\neq0$.

Additionally, $I_{\mathrm{corr}}$ reveals information on the spatial character of the noise spectrum. Assuming uniform qubit frequencies, $\omega_\alpha\equiv\omega_0$,
to simplify the analysis, we find
\begin{align}
    I_{\mathrm{corr}}(t)&=2\omega_0\sum_{\alpha\neq\beta}Q_{\alpha\beta}(t)\langle\Pi_\alpha^\dagger\Pi_\beta\rangle+I^{(\mathrm{ns})}_{\mathrm{corr}}(t),\label{eq:icor}\\
    Q_{\alpha\beta}(t)&=\int_{-\infty}^\infty\frac{\textrm d\omega}{2\pi}\bigg[S_{\alpha\beta}(\omega)\frac{\sin(\omega-\omega_0)t}{\omega-\omega_0}\nonumber\\ & \hspace{5em} -S_{\beta\alpha}(\omega)\frac{\sin(\omega+\omega_0)t}{\omega+\omega_0}\bigg],
\end{align}
where $I^{(\mathrm{ns})}_{\mathrm{corr}}(t)=2\omega_0\sum_{\alpha\neq\beta}Q^{(\mathrm{ns})}_{\alpha\beta}(t)\langle\Pi^\dagger_\alpha\Pi^\dagger_\beta\rangle+\mathrm{h.c.}$ is the nonsecular contribution to the correlated intensity, with h.c.\ the Hermitian conjugate. To reveal the meaning of $Q_{\alpha\beta}(t)$, let us use the Markovian approximation $\sin(xt)/x\approx\pi\delta(x)$, consequently $Q_{\alpha\beta}=[S_{\alpha\beta}(\omega_0)-S_{\beta\alpha}(-\omega_0)]/2$, which is the antisymmetrized noise power spectrum at $\omega_0$~\cite{bosco1}. From the partial energy of the system, $W_k=\sum_{\alpha=1}^k\omega_\alpha\langle Z_\alpha\rangle/2$, where $k<N$, we can calculate the correlated partial intensities $I^{(k)}_{\mathrm{corr}}$, that is, Eq.~\eqref{eq:icor} with the index $\alpha$ only going over the qubits $1\dots k$. It is the scaling of $I^{(k)}_{\mathrm{corr}}$ with increasing $k$ that carries the information on the spatial character of the noise ($\xi_{\alpha\beta}$). 
On the one hand, with a bath correlation length much longer than the register's size, the partial intensity $I^{(k)}_{\mathrm{corr}}$ increases with increasing $k\leq N$, as is also shown in Fig.~\ref{fig:1}b. On the other hand, when the correlation length includes only a subset of qubits (in this example it is limited to the first three qubits) the partial intensity stops increasing at $k=3$ (Fig.~\ref{fig:1}c).


\emph{Correlated dephasing---}We now consider longitudinal coupling to the environment, where the system operator appearing in Eq.~\eqref{eq:int} is $A_{\alpha,\varphi}=Z_\alpha$. This coupling leads to the fluctuation of the qubit frequencies, and hence to dephasing. In the absence of driving, the dissipator reads
\begin{equation}
\label{eq:dephase}
\begin{split}
\tilde D_{\varphi}&[t,\tilde\rho(t)]=-i\left[H_{ZZ}(t),\tilde\rho(t)\right]\\
&+\sum_{\alpha,\beta}\gamma_{\alpha\beta}^{(\varphi)}(t)\left[Z_\beta\tilde\rho(t)Z_\alpha-\frac{1}{2}\{Z_\alpha Z_\beta,\tilde\rho(t)\}\right],
\end{split}
\end{equation}
where the first term describes the bath-mediated $ZZ$ couplings between the qubits~\cite{supp}, while the second term entails correlated dephasing with time-dependent rates $\gamma_{\alpha\beta}^{(\varphi)}(t)=\int_{-\infty}^\infty\textrm d\omega/\pi\ S_{\alpha\beta,\varphi}(\omega)\sin(\omega t)/\omega$. This dissipator does not contribute to the intensity in Eq.~\eqref{eq:intensity1}, hence correlated dephasing does not interfere with the correlated relaxation discussed previously.

Considering only the effects from dephasing [Eq.~\eqref{eq:dephase}], the master equation [Eq.~\eqref{eq:TCL2}] turns into decoupled differential equations for the components of the density matrix in the computational basis (the eigenbasis of $H_0$). Particularly interesting are the equations for the anti-diagonal matrix elements $\tilde\rho_{\bar l,l}(t)$, where $l$ is a bitstring of length $N$ representing the state $|l\rangle$ and $\bar l$ its complement. These equations read
\begin{gather}
    \dot{\tilde\rho}_{\bar l, l}(t)=-2\tilde\rho_{\bar l,l}(t)\sum_{\alpha,\beta}(-1)^{l_\alpha+{l_\beta}}\gamma_{\alpha\beta}^{(\varphi)}(t),\label{eq:antidiag}
\end{gather}
with $l_\alpha\in\{0,1\}$ the $\alpha$th bit of $l$. In the absence of correlations, i.e., $\gamma_{\alpha\beta}^{(\varphi)}(t)=\delta_{\alpha\beta}\gamma_{\alpha}^{(\varphi)}(t)$, the dephasing rate of $\tilde\rho_{\bar l,l}(t)$ is the same for all $l$ and equals the sum of the local dephasing rates. 

\begin{figure}[t]
\includegraphics[width=\linewidth]{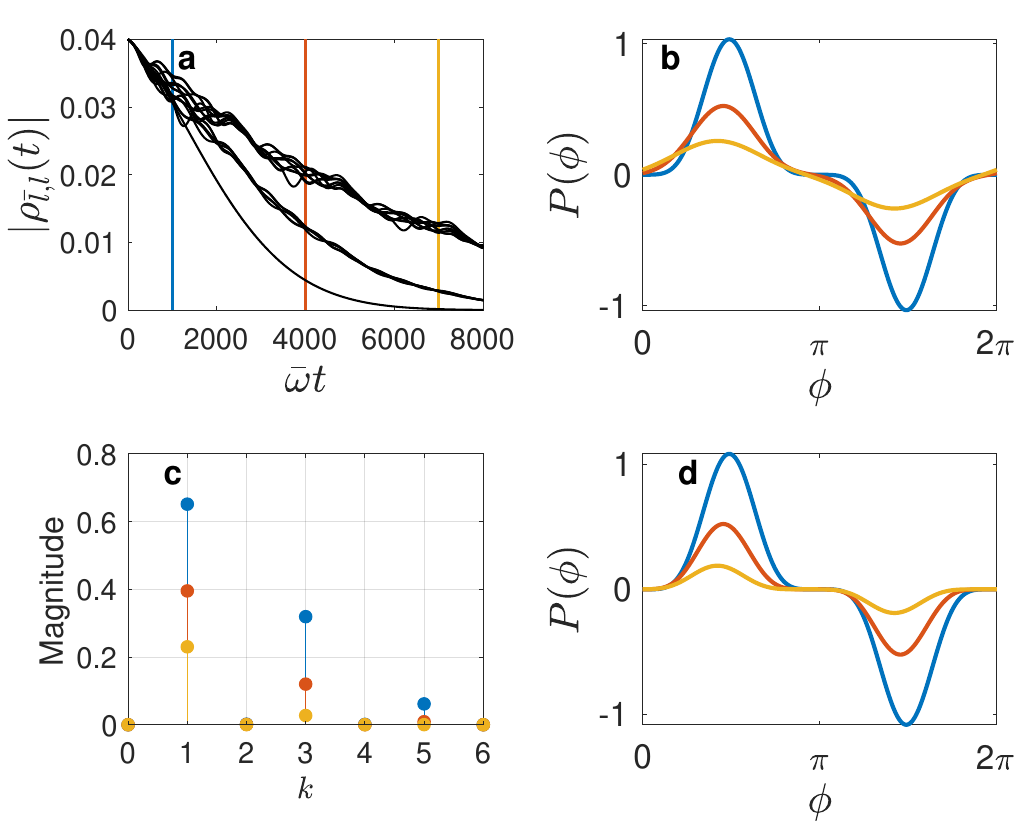}
\caption{\label{fig:2} Five-qubit system with nonuniform frequencies under correlated $1/f$ dephasing and Ohmic relaxation. The parameters are as in Fig.~\ref{fig:1} for relaxation and the strength of dephasing is $\lambda_\varphi/\bar \omega=10^{-9}$ with infrared cutoff $\omega_\mathrm{ir}/\bar\omega=10^{-6}$. (\textbf{a}) Separation of timescales in the dynamics of the anti-diagonal density matrix elements due to the longitudinally coupled environmental correlations. Each line represents one of 16 choices for $\{\bar l,l\}$. 
(\textbf{b}) Parity oscillation signal from the equal-superposition initial state after the different idling times that are indicated by the vertical lines in Fig.~\ref{fig:2}a. (\textbf{c}) The Fourier coefficients of the parity oscillations of Fig.~\ref{fig:2}b.  (\textbf{d}) Parity oscillations with uncorrelated dephasing, i.e., $\xi^{(\varphi)}_{\alpha\beta}\sim\delta_{\alpha\beta}$, showing unchanged signal shape, indicating the absence of spatial correlations in the longitudinally coupled noise.}
\end{figure}

According to Eq.~\eqref{eq:antidiag}, correlations lead to the separation of the dephasing rates for $\tilde\rho_{\bar l,l}(t)$ depending on $k$, which we define as the difference between the number of 1s and 0s in $l$.
For instance, the dephasing rate of the far off-diagonal matrix element $\trod$, representing the coherence between the ground state $|0\dots0\rangle$ and the fully inverted state $|1\dots1\rangle$, is $2\sum_{\alpha,\beta}\gamma_{\alpha\beta}^{(\varphi)}(t)$. Note that $\gamma_{\alpha\beta}^{(\varphi)}(t)\sim \xi^{(\varphi)}_{\alpha\beta}$, so that  $\sum_{\alpha,\beta}\gamma_{\alpha\beta}^{(\varphi)}(t)\sim N^2$ for fully correlated dephasing, as opposed to $\sum_{\alpha,\beta}\gamma_{\alpha\beta}^{(\varphi)}(t)\sim N$ for fully independent dephasing. This enhanced sensitivity of coherence, caused by noise correlations, is known as superdecoherence~\cite{palma1996quantum,Kattemolle2020conditions}. In contrast, the dephasing rates of the matrix elements with the same number of 1s and 0s in $l$ ($k=0$) are maximally reduced by the correlations. For a perfectly correlated environment ($\xi^{(\varphi)}_{\alpha\beta}=1$), these rates vanish, leading to the formation of a decoherence-free subspace~\cite{DFS}. If, in addition to correlated dephasing, there are (correlated) baths that couple transversely to the qubits, leading to dissipators in the form of Eq.~\eqref{eq:relax}, superdecoherence persists  due to the independent bath assumption~\cite{supp}. Also the separation of times over which the anti-diagonal matrix elements vanish persists, see Fig.~\ref{fig:2}a. 

Thus, the dynamics of the anti-diagonal elements reveal the presence or absence of correlated dephasing. To this end, we cluster the anti-diagonal elements of $\rho$ by $k$, and define $\rho^{(k)}$ as the sum of all matrix elements $\rho_{\bar l, l}$ such that the difference between the number of 1s and 0s in $l$ is $k$. A well-established method for obtaining the absolute value of $\rho^{(N)}(t)=\rod$ is by parity oscillations~\cite{GHZ,freeman1998spin}. Nevertheless, we show that parity oscillations contain low-frequency corrections arising from elements $\rho^{(k)}(t)$ ($k\neq N$).  In fact, all $\rho^{(k)}(t)$ can be extracted from these parity oscillations.  

Parity oscillations~\cite{GHZ,freeman1998spin}, here with generalized initial states, are obtained as follows. (1) Prepare an initial state $\rho(0)$ that populates the anti-diagonal matrix elements of interest. Possible choices include the equal-superposition state $\ket +^{\otimes N}$ and the $N$-qubit Greenberger--Horne--Zeilinger (GHZ) state $(\ket 0^{\otimes N}+\ket 1^{\otimes N})/\sqrt{2}$. (2) Wait for time $t$. (3) Apply the single-qubit gate $\mathrm{RZ}(\phi)\sqrt{X}^\dagger\,\mathrm{RZ}^\dagger(\phi)$ to each qubit and measure all qubits in the $Z$-basis.

The `parity', obtained from repetition of the above process, is defined as $P(\phi)=P_{\mathrm{even}}(\phi)-P_{\mathrm{odd}}(\phi)$, with $P_{\mathrm{even}/\mathrm{odd}}$ the probability of measuring an even/odd number of $-1$s in step (3). 
A direct computation~\cite{supp} shows that
\begin{align}
 P(\phi)=\sum_{k=-N}^{N} e^{i (\phi-\pi/2)k} \rho^{(k)}\label{eq:pk},
\end{align}
so that 
$\rho^{(k)}=\frac{\mathcal F [P'](k)}{2N+1}$,
with $\mathcal F[P'](k)$ the Fourier transform of $P'$, evaluated at $k$, and  $P'(x)=P\left(\frac{2\pi}{2N+1}x+\frac{\pi}{2}\right)$. Under the assumption that the only nonzero anti-diagonal entries of $\rho(t)$ are the far-off diagonal elements, which is the case for a perfect GHZ state, $\lvert \rod \rvert$ is half the amplitude of $P(\phi)$. In experiments with high-fidelity GHZ states, where additionally dephasing  is the dominant noise channel, the above assumption is justified. However, in general, it is not, e.g., for noisy GHZ states, when relaxation cannot be ignored, or for states such as $\ket{+}^{\otimes N}$. 

Previous experimental evidence for correlated dephasing has been obtained by creating the GHZ state and observing the degradation of the far off-diagonal matrix element by parity oscillations \cite{GHZ,ozaeta2019decoherence}. The scaling behavior of this degradation with increasing qubit numbers demonstrates correlated dephasing through superdecoherence. However, it is challenging to generate high-fidelity many-qubit GHZ states; additionally, precisely due to superdecoherence, the oscillation amplitude quickly becomes very weak, making it difficult to monitor. 

Based on Eq.~\eqref{eq:pk}, a much simpler alternative approach is to generate parity oscillations from the equal superposition state, putting equal initial weights on the anti-diagonal elements. As the correlations lead to distinct decay times for these elements, the Fourier components of the parity oscillations diminish at different rates. The resulting change in $P(\phi)$ directly reveals the presence of correlations, as illustrated in Fig.~\ref{fig:2} using a five-qubit ($N=5$) example subjected to correlated $1/f$ dephasing ($S_{\alpha\beta,\varphi}(\omega)=\lambda_\varphi/|\omega|$) and Ohmic relaxation as discussed previously.

We note that `multiple quantum coherences' (MQC)~\cite{wei2020verifying,baum1985multiple,supp} offer an alternative to parity oscillations. However, they provide information about $\rho$ different from $\rho^{(k)}$. Furthermore, the validity of the method that relates the MQC signal to the information about $\rho$ has only been demonstrated under restricted noise channels, which are presupposed to have uncorrelated errors, among other limitations~\cite{gaerttner2018relating,supp}.

\emph{Conclusion---}We proposed simple methods that reveal the presence of spatially correlated noise in qubit registers. Initializing the system in the fully inverted state and then monitoring the dynamics of the derivative of the total energy shows the presence of correlated environments that couple transversely to the qubits. Moreover, observing the change in the line shape of parity oscillations from the initial equal-superposition state signals the presence of correlated dephasing. Both procedures require only single-qubit control, making them readily implementable on all currently available platforms, even before the quantum processor in question is fully calibrated or integrated. This leads to shorter design feedback loops, accelerating the development of quantum computing hardware. 

\begin{acknowledgments}
\emph{Acknowledgments---}We acknowledge the support from the German Ministry for Education and Research, under the QSolid project, Grant No.~13N16167, and from the State of Baden-W\"urttemberg within the Competence Center Quantum Computing, project KQCBW24.
\end{acknowledgments}

\bibliography{spacecorr}

\appendix
\widetext
\clearpage

\begin{center}
    {\huge\textbf{Supplemental material}}
\end{center}

\vspace{0.5cm}

{\Large\textbf{I.\quad Time-convolutionless formalism}}
\vspace{0.5cm}

\large The starting Hamiltonian of our closed system describes individually addressable qubits without couplings, their Hamiltonian in the lab frame is $(\hbar=1)$
\begin{equation}\label{Ham}
    H(t) = H_0 + H_{\textrm{ct}}(t) = \sum_{\alpha=1}^N \left(\frac{\omega_{\alpha}}{2}Z_\alpha+\Omega_\alpha(t)X_\alpha\right).
\end{equation}
Here, $\alpha$ indexes the different qubits, the number of qubits is $N$, the qubit frequencies are $\omega_{\alpha}$, the operators $Z_\alpha,X_\alpha$ are the standard Pauli operators acting on the $\alpha$-th qubit, and $\Omega_\alpha(t)$ is an optional pulse controlling the individual qubits. This system is coupled to multiple quantum environments with total Hamiltonian $H_B$. We assume the environment coupling is linear and weak compared to the other energy scales of the system, hence the microscopic details of the environments are irrelevant and we only require the bath correlation functions. The interactions between system and the baths are
\begin{equation}
    H_{\textrm{SB}} = \sum_{\alpha,j} A_{\alpha,j} \otimes B_{\alpha,j},
\end{equation}
where the index $j$ is an effect index, e.g. local relaxation or correlated dephasing. The environment couples with the Hermitian bath operator $B_{\alpha,j}$ to the system via $A_{\alpha,j}$. We emphasize again, in the weak coupling description the dissipative effects depend only on the spectral properties of the environment, which is described in its correlation functions and their Fourier transforms,
\begin{equation}
    C_{\alpha\beta;jk}(t)=\langle B_{\alpha,j}(t)B_{\beta,k}(0)\rangle=\delta_{jk}\int_{-\infty}^\infty \frac{\textrm d\omega}{2\pi}\  S_{\alpha\beta;j}(\omega)e^{-i\omega t}.\label{spectra}
\end{equation}

{\large\textbf{A.\quad Closed system dynamics}}
\vspace{0.5cm}

Due to the weak coupling assumption, the environment affects the qubit hardware perturbatively. As such, we first require a description for the unperturbed dynamics. This is described by the unitary propagator, which reads in the lab frame $\rho(t)=U(t,t_0)\rho(t_0)U^\dagger(t,t_0)$, where $\rho(t)$ is the state of the qubits in the lab frame. To separate the trivial dynamics from the controlled one, we introduce the interaction frame
in which the density matrix of the system is $\tilde\rho(t)=e^{iH_0t}\rho(t)e^{-iH_0t}$. 
The relation between interaction frame and lab frame propagator is $U(t,t_0)=e^{-iH_0t}\tilde U(t,t_0)e^{iH_0t_0}$,
the interaction frame propagator is obtained as the solution of the following Schrödinger equation, $i\partial_t\tilde U(t,t_0)=\tilde H_{\textrm{ct}}(t)\tilde U(t,t_0)$, with the interaction frame control $\tilde H_{\textrm{ct}}(t)=e^{iH_0t}H_{\textrm{ct}}(t)e^{-iH_0t}$ and initial condition $\tilde U(t_0,t_0)=\mathbb 1$. The useful properties of the propagator are its divisibility and inverse, which read in the lab frame: (1) $U(t,t_0)=U(t,t_1)U(t_1,t_0)$, where $(t>t_1>t_0)$, (2) $U^\dagger(t,t')=U(t',t)$.
These properties are inherited by the interaction frame propagator $\tilde U$, which is confirmed by direct substitutions. With this separation, we can model general driven systems and take $\tilde U=\mathbb 1$ for the undriven examples. The initial time $t_0$ is chosen as $t_0=0$ without loss of generality and we abbreviate $\tilde U(t,t_0)$ with $\tilde U(t)$.

\vspace{0.5cm}
{\large\textbf{B.\quad Interaction picture}}
\vspace{0.5cm}

As the system interacts with a time-dependent field and a collection of environments simultaneously, we may have two `interaction pictures'. For the control field, we call it the interaction frame and we use the tilde notation for its indication. The interaction picture is reserved with respect to the environment, and it is this picture within which the time-convolutionless master equation can be derived. The interaction Hamiltonian in this picture reads
\begin{equation}
    H_I(t)=\sum_{\alpha,j}A_{\alpha,j}(t)B_{\alpha,j}(t), 
\end{equation}
with the operators being
\begin{align}
    A_{\alpha,j}(t)=\tilde U^\dagger(t)e^{iH_0t}A_{\alpha,j}e^{-iH_0t}\tilde U(t)=\tilde U^\dagger(t)\tilde A_{\alpha,j}(t)\tilde U(t),\\
    B_{\alpha,j}(t)=e^{iH_Bt}B_{\alpha,j}e^{-iH_Bt}. 
\end{align}
The combined system-environment state in the interaction picture is
\begin{equation}
    \rho_I(t)=\tilde U^\dagger(t)e^{iH_0t}e^{iH_Bt}\rho_{SE}(t)e^{-iH_Bt}e^{-iH_0t}\tilde U(t).
\end{equation}
After tracing out the bath, the reduced state in the interaction picture is
\begin{align}
    \rho_{I,S}(t)=\tilde U^\dagger(t)e^{iH_0t}\textrm{Tr}_B\left(e^{iH_Bt}\rho_{SE}(t)e^{-iH_Bt}\right)e^{-iH_0t}\tilde U(t)\nonumber\\
    =\tilde U^\dagger(t)e^{iH_0t}\rho(t)e^{-iH_0t}\tilde U(t)=\tilde U^\dagger(t)\tilde \rho(t)\tilde U(t),
\end{align}
where $\rho(t)$ is the lab frame state of the reduced system and $\tilde\rho(t)$ is the interaction frame state of the reduced system. Setting $\tilde U=\mathbb 1$ gives the conventional undriven special cases immediately for all the above formulas.

\vspace{0.5cm}
{\large\textbf{C.\quad Time-convolutionless master equation}}
\vspace{0.5cm}

The second-order time-convolutionless master equation (TCL2) for the state of the qubits in the interaction picture is \cite{breuer2002theory}
\begin{equation}
    \frac{\textrm d}{\textrm dt}\rho_{I,S}(t) = -\int_0^t\textrm ds\ \textrm{Tr}_B\big[H_I(t),[H_I(s),\rho_{I,S}(t)\otimes\rho_B]\big].\label{tcl2}
\end{equation}
The assumptions are the \emph{Born} approximation, expressing the weak coupling between the system and its environment, and an initially factorizing system-bath state, $\rho_I(0)=\rho_{I,S}(0)\otimes\rho_B$, with the environmental reference state $\rho_B$ being an equilibrium state, $[H_B,\rho_B]=0$. We note that in the presence of multiple environments, $\rho_B$ is a tensor product of individual states, each of which describes the equilibrium state of its corresponding environment. As the bath Hamiltonian $H_B$ does not contain interaction terms between the different environments, the correlation functions in Eq.~\eqref{spectra} are indeed uncorrelated across the different baths \cite{Addgen}.

After straightforward substitutions, we first write the TCL2 equation in the lab frame to see its structure,
\begin{equation}
    \frac{\textrm d}{\textrm dt}\rho_{}(t) = -i[H(t),\rho(t)]+\sum_jD_j(t,\rho).
\end{equation}
As always, we have the Hamiltonian component of the dynamical equation with the full time-dependent Hamiltonian from Eq.~\eqref{Ham}. The effects of the environments are described by the independent dissipators indexed by $j$, these are
\begin{align}
     D_j(t,\rho)=\sum_{\alpha,\beta}\int_0^t\textrm ds\ C_{\alpha\beta;j}(t-s)\left(U(t,s)A_{\beta,j}U(s,t)\rho(t)A_{\alpha,j}-A_{\alpha,j}U(t,s)A_{\beta,j}U(s,t)\rho(t)\right)\nonumber\\
     +\sum_{\alpha,\beta}\int_0^t\textrm ds\ C_{\beta\alpha;j}(s-t)\left(A_{\alpha,j}\rho(t)U(t,s)A_{\beta,j}U(s,t)-\rho(t)U(t,s)A_{\beta,j}U(s,t)A_{\alpha,j}\right).
\end{align}
Generally, the spatial correlations appear in the dissipator which cause couplings between the qubits, hence their dynamics are not independent. Furthermore, the presence of the full unitary in the formulas indicates that the effects of the time-dependent fields influence the way the environment acts on the system. 

To proceed, we transform the TCL2 equation into the interaction frame to arrive at Eq.~\eqref{eq:TCL2} of the main text,
\begin{equation}
    \frac{\textrm d}{\textrm dt}\tilde \rho_{}(t) = -i[\tilde H_{\textrm{ct}}(t),\tilde\rho(t)]+\sum_j\tilde D_j(t,\tilde\rho),\label{maintext3}
\end{equation}
with the dissipators being
\begin{align}
     \tilde D_j=\sum_{\alpha,\beta}\int_0^t\textrm ds\ C_{\alpha\beta;j}(t-s)\left(\tilde U(t,s)\tilde A_{\beta,j}(s)\tilde U(s,t)\tilde\rho(t)\tilde A_{\alpha,j}(t)-\tilde A_{\alpha,j}(t)\tilde U(t,s)\tilde A_{\beta,j}(s)\tilde U(s,t)\tilde\rho(t)\right)\nonumber\\
     +\sum_{\alpha,\beta}\int_0^t\textrm ds\ C_{\beta\alpha;j}(s-t)\left(\tilde A_{\alpha,j}(t)\tilde\rho(t)\tilde U(t,s)\tilde A_{\beta,j}(s)\tilde U(s,t)-\tilde\rho(t)\tilde U(t,s)\tilde A_{\beta,j}(s)\tilde U(s,t)\tilde A_{\alpha,j}(t)\right).\label{drivendiss}
\end{align}
In the absence of control fields, the dynamics in the interaction frame is dictated only by the dissipator, since $\tilde H_{\textrm{ct}}(t)=0$ and $\tilde U(t,s)=\mathbb 1$. 

\vspace{0.5cm}
{\large\textbf{D.\quad Frequency domain dissipator}}
\vspace{0.5cm}

We can interpret the dissipator very well in the frequency domain. The interaction frame operators can be written as 
\begin{equation}
    \tilde A_{\alpha,j}(t)=e^{iH_\alpha t}A_{\alpha,j}e^{-iH_\alpha t}=\sum_{\omega}e^{i\omega t}A_{\alpha,j}(\omega)=\sum_{\omega}e^{-i\omega t}A^\dagger_{\alpha,j}(\omega),\label{freq1}
\end{equation}
where $\omega$ is all the possible Bohr frequencies of $H_\alpha$ and $A_{\alpha,j}(\omega)$ are the eigenoperators of $H_\alpha$, the last equality expresses Hermiticity. 
We now substitute the expression of the correlation function from Eq.~\eqref{spectra} and the eigenoperator decomposition from Eq.~\eqref{freq1} to Eq.~\eqref{drivendiss} in the absence of drives, i.e., $\tilde U(t,s)=\mathbb 1$, we obtain 
\begin{align}
\tilde D_j(t,\tilde\rho)=\sum_{\substack{\alpha,\beta \\ \omega',\omega''}}\Gamma_{\alpha\beta,j}(\omega',\omega'',t)\left(A_{\beta,j}(\omega')\tilde\rho(t)A^\dagger_{\alpha,j}(\omega'')- A^\dagger_{\alpha,j}(\omega'')A_{\beta,j}(\omega')\tilde\rho(t)\right)+\mathrm{h.c.}\label{ME}
\end{align}
Here, we introduced the rates
\begin{gather}
    \Gamma_{\alpha\beta,j}(\omega',\omega'',t)=\int_{-\infty}^\infty \frac{\textrm d\omega}{2\pi}\  S_{\alpha\beta;j}(\omega)e^{-i(\omega+\omega'') t} \int_0^t\textrm ds\ e^{i(\omega+\omega')s}\nonumber\\
    =e^{i(\omega'-\omega'')t}\int_{-\infty}^\infty \frac{\textrm d\omega}{2\pi }\  S_{\alpha\beta;j}(\omega)\frac{1-e^{-i(\omega+\omega')t}}{i(\omega+\omega')}.\label{rates}
\end{gather}
We must bear in mind that here $\omega'$ and $\omega''$ are discrete variables associated with the Bohr frequencies. 
Inside the frequency integral, the filter function, 
\begin{equation}
    F(\Omega,t)=\frac{1-e^{-i\Omega t}}{i\Omega}\label{eq:filt}
\end{equation}
appears, so that
\begin{equation}
    \Gamma_{\alpha\beta,j}(\omega',\omega'',t)=e^{i(\omega'-\omega'')t}\int_{-\infty}^\infty \frac{\textrm d\omega}{2\pi}\  S_{\alpha\beta;j}(\omega)F(\omega+\omega',t).\label{rata}
\end{equation}
Equations~\eqref{ME} and \eqref{rata} are the starting points for the discussion of the correlated phenomena for the qubit register.

\vspace{0.5cm}
{\Large\textbf{II.\quad Correlated relaxation}}
\vspace{0.5cm}

As discussed in the main text, the system operator appearing in the master equation is $A_{\alpha,\mathrm{relax}}=X_\alpha$.
The eigenoperator decomposition of this operator in Eq.~\eqref{freq1} is $X_\alpha=e^{-i\omega_\alpha t}\Pi_\alpha+e^{i\omega_\alpha t}\Pi_\alpha^\dagger$. The local raising and lowering operators act on the qubit states as, $\Pi_\alpha|0\rangle_\alpha=0$, $\Pi_\alpha|1\rangle_\alpha=|0\rangle_\alpha $, $\Pi^\dagger_\alpha|0\rangle_\alpha=|1\rangle_\alpha $, $\Pi^\dagger_\alpha|1\rangle_\alpha=0$. Thus, the eigenoperators appearing in the dissipator Eq.~\eqref{ME} for correlated relaxation with the corresponding frequencies are
\begin{equation}
    A_{\alpha,\mathrm{relax}}(\omega)=\begin{cases}
\Pi_\alpha,\quad \omega=-\omega_{\alpha},\\
\Pi^\dagger_\alpha,\quad \omega=\omega_{\alpha}.
\end{cases}
\end{equation}
This leads to the dissipator
\begin{align}
\tilde D_{\textrm{relax}}(t,\tilde\rho)=\sum_{\alpha,\beta}\Gamma_{\alpha\beta}(\omega_{\beta},\omega_{\alpha},t)\left(\Pi^\dagger_{\beta}\tilde\rho(t)\Pi_{\alpha}- \Pi_{\alpha}\Pi^\dagger_{\beta}\tilde\rho(t)\right)\nonumber\\
+\sum_{\alpha,\beta}\Gamma_{\alpha\beta}(-\omega_{\beta},-\omega_{\alpha},t)\left(\Pi_{\beta}\tilde\rho(t)\Pi^\dagger_{\alpha}- \Pi^\dagger_{\alpha}\Pi_{\beta}\tilde\rho(t)\right)\nonumber\\
+\sum_{\alpha,\beta}\Gamma_{\alpha\beta}(\omega_{\beta},-\omega_{\alpha},t)\left( \Pi^\dagger_{\beta}\tilde\rho(t)\Pi^\dagger_{\alpha}-\Pi^\dagger_{\alpha}\Pi^\dagger_{\beta}\tilde\rho(t)\right)\nonumber\\
+\sum_{\alpha,\beta}\Gamma_{\alpha\beta}(-\omega_{\beta},\omega_{\alpha},t)\left( \Pi_{\beta}\tilde\rho(t)\Pi_{\alpha}-\Pi_{\alpha}\Pi_{\beta}\tilde\rho(t)\right)+\mathrm{h.c.}\label{eq:relaxsupp}
\end{align}
The first two lines here correspond to the secular terms and the last two terms are the nonsecular terms, Hermitian conjugation is meant for all terms. According to Eq.~\eqref{rata} when the qubit frequencies appear with different signs, the exponential will rapidly oscillate. If the integral does not cancel these rapid oscillations then the so-called nonsecular terms are neglected under the secular approximation, hence the terminology. In the rates the index $j$ corresponds to ``relax" and it is omitted.

Now, we combine the expressions with their Hermitian conjugate. This reads for the secular terms,
\begin{align}
\tilde D^{\textrm{(sec)}}_{\textrm{relax}}=-i\left[\sum_{\alpha,\beta}\frac{\Gamma_{\alpha\beta}(\omega_{\beta},\omega_{\alpha},t)-\Gamma_{\beta\alpha}^*(\omega_{\alpha},\omega_{\beta},t)}{2i}\Pi_\alpha \Pi^\dagger_\beta,\tilde\rho(t)\right]\nonumber\\
-i\left[\sum_{\alpha,\beta}\frac{\Gamma_{\alpha\beta}(-\omega_{\beta},-\omega_{\alpha},t)-\Gamma_{\beta\alpha}^*(-\omega_{\alpha},-\omega_{\beta},t)}{2i}\Pi^\dagger_\alpha \Pi_\beta,\tilde\rho(t)\right]\nonumber\\
+\sum_{\alpha,\beta}\left(\Gamma_{\alpha\beta}(\omega_{\beta},\omega_{\alpha},t)+\Gamma_{\beta\alpha}^*(\omega_{\alpha},\omega_{\beta},t)\right)\left(\Pi^\dagger_\beta\tilde\rho(t)\Pi_\alpha-\frac{1}{2}\{\Pi_\alpha \Pi^\dagger_\beta,\tilde\rho(t)\}\right)\nonumber\\
+\sum_{\alpha,\beta}\left(\Gamma_{\alpha\beta}(-\omega_{\beta},-\omega_{\alpha},t)+\Gamma_{\beta\alpha}^*(-\omega_{\alpha},-\omega_{\beta},t)\right)\left(\Pi_\beta\tilde\rho(t)\Pi^\dagger_\alpha-\frac{1}{2}\{\Pi^\dagger_\alpha \Pi_\beta,\tilde\rho(t)\}\right),
\end{align}
and the nonsecular terms,
\begin{align}
\tilde D^{\textrm{(nonsec)}}_{\textrm{relax}}=-i\left[\sum_{\alpha,\beta}\frac{\Gamma_{\alpha\beta}(\omega_{\beta},-\omega_{\alpha},t)-\Gamma_{\beta\alpha}^*(-\omega_{\alpha},\omega_{\beta},t)}{2i}\Pi^\dagger_\alpha \Pi^\dagger_\beta,\tilde\rho(t)\right]\nonumber\\
-i\left[\sum_{\alpha,\beta}\frac{\Gamma_{\alpha\beta}(-\omega_{\beta},\omega_{\alpha},t)-\Gamma_{\beta\alpha}^*(\omega_{\alpha},-\omega_{\beta},t)}{2i}\Pi_\alpha \Pi_\beta,\tilde\rho(t)\right]\nonumber\\
+\sum_{\alpha,\beta}\left(\Gamma_{\alpha\beta}(\omega_{\beta},-\omega_{\alpha},t)+\Gamma_{\beta\alpha}^*(-\omega_{\alpha},\omega_{\beta},t)\right)\left(\Pi^\dagger_\beta\tilde\rho(t)\Pi^\dagger_\alpha-\frac{1}{2}\{\Pi^\dagger_\alpha \Pi^\dagger_\beta,\tilde\rho(t)\}\right)\nonumber\\
+\sum_{\alpha,\beta}\left(\Gamma_{\alpha\beta}(-\omega_{\beta},\omega_{\alpha},t)+\Gamma_{\beta\alpha}^*(\omega_{\alpha},-\omega_{\beta},t)\right)\left(\Pi_\beta\tilde\rho(t)\Pi_\alpha-\frac{1}{2}\{\Pi_\alpha \Pi_\beta,\tilde\rho(t)\}\right).
\end{align}

\vspace{0.5cm}
{\large\textbf{A.\quad Compact form of the relaxation dissipator}}
\vspace{0.5cm}

We now rewrite the entire relaxation dissipator (secular and nonsecular terms) into the form of Eq.~\eqref{eq:relax} of the main text,
\begin{align}
\tilde D_{\textrm{relax}}(t,\tilde\rho)=-i\left[H_{XY}(t),\tilde\rho(t)\right]
+\sum_{\substack{\alpha,\beta \\ i,j}}\gamma_{\alpha\beta}^{(ij)}(t)\left(\Pi^{(i)}_\beta\tilde\rho(t)\Pi^{(j)}_\alpha-\frac{1}{2}\{\Pi^{(j)}_\alpha \Pi^{(i)}_\beta,\tilde\rho(t)\}\right),\label{relax2}
\end{align}
where $\{i,j\}\in\{1,2\}$ and we introduced the notation $\Pi^{(1)}=\Pi^\dagger,\ \Pi^{(2)}=\Pi$. In the following, we will also use the vector of Pauli matrices $\vec P=(X\ Y\ Z)^T$. The Hamiltonian contribution in the relaxation dissipator is
\begin{align}
    H_{XY}(t)=\sum_{\alpha,\beta}J_{\alpha\beta}^{XX}(t)X_\alpha X_\beta+J_{\alpha\beta}^{YY}(t)Y_\alpha Y_\beta+i\mathcal D_{\alpha\beta}(t)\hat z\cdot(\vec P_\alpha\times\vec P_\beta)\nonumber\\
    +\sum_{\alpha,\beta}J_{\alpha\beta}^{XY}(t)\big(X_\alpha Y_\beta+Y_\alpha X_\beta\big),\label{hxy}
\end{align}
with time-dependent exchange interaction strengths
\begin{align}
    J_{\alpha\beta}^{XX}(t)=\frac{1}{4}\left(J_{\alpha\beta}^{(1)}(t)+J_{\alpha\beta}^{(2)}(t)+2\textrm{Re}\big(J_{\alpha\beta}^{(3)}(t)\big)\right),\\
    J_{\alpha\beta}^{YY}(t)=\frac{1}{4}\left(J_{\alpha\beta}^{(1)}(t)+J_{\alpha\beta}^{(2)}(t)-2\textrm{Re}\big(J_{\alpha\beta}^{(3)}(t)\big)\right),\\
    \mathcal D_{\alpha\beta}(t)=\frac{1}{4}\left(J_{\alpha\beta}^{(1)}(t)-J_{\alpha\beta}^{(2)}(t)\right),\\
    J_{\alpha\beta}^{XY}(t)=-\frac{1}{2}\textrm{Im}\big(J_{\alpha\beta}^{(3)}(t)\big).
\end{align}
Here, we see the linear combination of the following integrals,
\begin{align}
    J_{\alpha\beta}^{(1)}(t)=e^{i(\omega_\beta-\omega_\alpha)t}\int_{-\infty}^\infty \frac{\textrm d\omega}{4\pi i}\  S_{\alpha\beta}(\omega)\big(F(\omega+\omega_\beta,t)-F^*(\omega+\omega_\alpha,t)\big),\\
    J_{\alpha\beta}^{(2)}(t)=e^{i(\omega_\alpha-\omega_\beta)t}\int_{-\infty}^\infty \frac{\textrm d\omega}{4\pi i}\  S_{\alpha\beta}(\omega)\big(F(\omega-\omega_\beta,t)-F^*(\omega-\omega_\alpha,t)\big),\\
    J_{\alpha\beta}^{(3)}(t)=e^{i(\omega_\beta+\omega_\alpha)t}\int_{-\infty}^\infty \frac{\textrm d\omega}{4\pi i}\  S_{\alpha\beta}(\omega)\big(F(\omega+\omega_\beta,t)-F^*(\omega-\omega_\alpha,t)\big).
\end{align}
In the Hamiltonian, we have symmetric exchange interactions $XX$ and $YY$ with the nonsecular term $J^{(3)}_{\alpha\beta}(t)$ introducing anisotropy. The third term is an antisymmetric exchange interaction, i.e., Dzyaloshinskii–Moriya interaction, and the last term originating from the nonsecular terms describing another symmetric exchange interaction.

The time-dependent decay rates appearing in the relaxation dissipator are
\begin{align}
    \gamma_{\alpha\beta}^{(12)}(t)=e^{i(\omega_\beta-\omega_\alpha)t}\int_{-\infty}^\infty \frac{\textrm d\omega}{2\pi}\  S_{\alpha\beta}(\omega)\big(F(\omega+\omega_\beta,t)+F^*(\omega+\omega_\alpha,t)\big),\label{eq:rates1}\\
    \gamma_{\alpha\beta}^{(21)}(t)=e^{i(\omega_\alpha-\omega_\beta)t}\int_{-\infty}^\infty \frac{\textrm d\omega}{2\pi}\  S_{\alpha\beta}(\omega)\big(F(\omega-\omega_\beta,t)+F^*(\omega-\omega_\alpha,t)\big),\label{eq:rates2}\\
    \gamma_{\alpha\beta}^{(11)}(t)=e^{i(\omega_\beta+\omega_\alpha)t}\int_{-\infty}^\infty \frac{\textrm d\omega}{2\pi}\  S_{\alpha\beta}(\omega)\big(F(\omega+\omega_\beta,t)+F^*(\omega-\omega_\alpha,t)\big).\label{eq:rates3}
\end{align}
Here, the filter functions appear as linear combinations of Eq.~\eqref{eq:filt} evaluated at frequencies related to the particular qubits. In the main text, Eqs.~\eqref{eq:rates1}--\eqref{eq:rates3} are abbreviated to $\gamma_{\alpha\beta}^{(ij)}(t)=\int_{-\infty}^{\infty}\mathrm{d}\omega/2\pi\ S_{\alpha\beta}(\omega)F_{\alpha\beta}^{(ij)}(\omega,t)$.

\vspace{0.5cm}
{\Large\textbf{III.\quad Correlated dephasing}}
\vspace{0.5cm}

For dephasing processes, the coupling to the environment is longitudinal, $A_{\alpha,\varphi}=Z_\alpha$. The eigenoperator and its frequency for the dissipator in Eq.~\eqref{ME} is
\begin{equation}
    A_{\alpha,j}(\omega)=Z_\alpha,\quad \omega=0.
\end{equation}
This leads to the dephasing dissipator
\begin{align}
\tilde D_{\varphi}(t,\tilde \rho)=\sum_{\alpha,\beta}\Gamma_{\alpha\beta,\varphi}(t)\left[Z_\beta\tilde\rho(t)Z_\alpha-Z_\alpha Z_\beta\tilde\rho(t)\right]+\mathrm{h.c.}.
\end{align}
We can separate a Hamiltonian contribution from the dissipators after some algebraic manipulations of the terms and their Hermitian conjugates, yielding
\begin{align}
\tilde D_{\varphi}(t,\tilde \rho)=-i\left[\sum_{\alpha,\beta}\frac{\Gamma_{\alpha\beta,\varphi}(t)-\Gamma_{\beta\alpha,\varphi}^*(t)}{2i}Z_\alpha Z_\beta,\tilde\rho(t)\right]\nonumber\\
+\sum_{\alpha,\beta}\left[\Gamma_{\alpha\beta,\varphi}(t)+\Gamma^*_{\beta\alpha,\varphi}(t)\right]\left(Z_\beta\tilde\rho(t)Z_\alpha-\frac{1}{2}\{Z_\alpha Z_\beta,\tilde\rho(t)\}\right).\label{dephase1}
\end{align}
The dephasing dissipator in Eq.~\eqref{dephase1} is rewritten into the form appearing as Eq.~\eqref{eq:dephase} in the main text, which reads
\begin{align}
\tilde D_{\varphi}(t,\tilde\rho)=-i\left[H_{ZZ}(t),\tilde\rho(t)\right]
+\sum_{\alpha,\beta}\gamma_{\alpha\beta}^{(\varphi)}(t)\left(Z_\beta\tilde\rho(t)Z_\alpha-\frac{1}{2}\{Z_\alpha Z_\beta,\tilde\rho(t)\}\right).\label{dephase2}
\end{align}
Here, 
\begin{equation}
H_{ZZ}(t)=\sum_{\alpha,\beta}J_{\alpha\beta}^{ZZ}(t)Z_\alpha Z_\beta,
\end{equation}
with
\begin{equation}
J_{\alpha\beta}^{ZZ}(t)=\int_{-\infty}^\infty \frac{\textrm d\omega}{2\pi}\  S_{\alpha\beta;\varphi}(\omega)\textrm{Im}\big[F(\omega,t)\big],
\end{equation}
and 
\begin{equation}
\gamma_{\alpha\beta}^{(\varphi)}(t)=\int_{-\infty}^\infty \frac{\textrm d\omega}{\pi}\  S_{\alpha\beta;\varphi}(\omega)\textrm{Re}\big[F(\omega,t)\big].
\end{equation}

\vspace{0.5cm}
{\large\textbf{A.\quad Equation of the far off-diagonal element and superdecoherence}}
\vspace{0.5cm}

Using the relaxation dissipator in Eq.~\eqref{relax2} along with the dephasing dissipator in Eq.~\eqref{dephase2}, in the absence of control terms the equation for the far off-diagonal matrix element can be calculated from
\begin{align}
    \langle 0\dots0|\frac{\mathrm d}{\mathrm dt}\tilde\rho(t)|1\dots1\rangle = \langle 0\dots0|\left(\tilde D_{\mathrm{relax}}(t,\tilde\rho)+\tilde D_{\varphi}(t,\tilde\rho)\right)|1\dots1\rangle,
\end{align}
which leads to
\begin{gather}
    \frac{\mathrm d}{\mathrm dt}\tilde\rho_{0\cdots0,1\cdots1}(t)=\left(-\frac{1}{2}\sum_\alpha\left(\gamma_{\alpha\alpha}^{\uparrow}(t)+\gamma_{\alpha\alpha}^{\downarrow}(t)\right)-2\sum_{\alpha,\beta}\gamma_{\alpha\beta}^{(\varphi)}(t)\right)\tilde\rho_{0\cdots0,1\cdots1}(t)\nonumber\\
    +\sum_{\alpha\neq\beta}\left(iJ_{\alpha\beta}^{(3)}(t)-\frac{\gamma^{(11)}_{\beta\alpha}(t)}{2}\right)^*\tilde\rho_{0\cdots1_\alpha\cdots1_\beta\cdots0,1\cdots1}(t)+\sum_{\alpha\neq\beta}\left(-iJ_{\alpha\beta}^{(3)}(t)-\frac{\gamma^{(11)}_{\beta\alpha}(t)}{2}\right)^*\tilde\rho_{0\cdots0,1\cdots0_\alpha\cdots0_\beta\cdots1}(t)\nonumber\\
    +\sum_{\alpha,\beta}\left(\gamma_{\beta\alpha}^{(11)}(t)\right)^*\tilde\rho_{0\cdots1_\beta\cdots0,1\cdots0_\alpha\cdots1}(t).
\end{gather}
Here, the subscripts indicate an entry at the particular position, e.g. $0\cdots1_\alpha\cdots1_\beta\cdots0$ means the bitstring has entries $0$, except at the positions $\alpha$ and $\beta$ where it has entries $1$. This differential equation shows that the far off-diagonal element only couples to other matrix elements through the nonsecular terms originating from the correlated relaxation dissipator. The equation's solution can be written as the sum of the solution of the homogeneous part of the equation and a particular solution of the inhomogeneous part, which reads
\begin{align}
    \tilde\rho_{0\cdots0,1\cdots1}(t)=\tilde\rho_{0\cdots0,1\cdots1}(0)\exp\left(\int_0^t\mathrm ds\ \left(-\frac{1}{2}\sum_\alpha\left(\gamma_{\alpha\alpha}^{\uparrow}(s)+\gamma_{\alpha\alpha}^{\downarrow}(s)\right)-2\sum_{\alpha,\beta}\gamma_{\alpha\beta}^{(\varphi)}(s)\right)\right)+\Delta_{\mathrm{ns}}(t).
\end{align}
The function appearing in the exponential, usually referred to as the decoherence function \cite{breuer2002theory}, shows that this matrix element has enhanced sensitivity only to the correlations arising from longitudinally coupled environments. Thus, superdecoherence persists in the presence of correlated environments that couple transversely to the qubits. The term $\Delta_{\mathrm{ns}}$ stemming from the nonsecular terms accumulates with increasing qubit numbers and may thus obscure superdecoherence.

\vspace{0.5cm}
{\large\textbf{B.\quad Parity oscillations}}
\vspace{0.5cm}


It is commonly stated that the sum of the absolute values of the two far off-diagonal matrix elements,
\begin{equation}
C=\lvert\rho_{0\cdots 0,1\cdots 1}\rvert+\lvert\rho_{1\cdots 1,0\cdots 0}\rvert=2\lvert \rho_{0\cdots 0, 1\cdots 0}\rvert,
\end{equation}
is equal to the amplitude of the parity oscillations \cite{ozaeta2019decoherence,GHZ,freeman1998spin}. 
These oscillations are obtained by first applying the local unitary
\begin{equation}
  U(\phi)=\frac{1}{\sqrt{2}}\left[\mathbb{1} + i
    \left(\begin{array}{cc}
      0 & e^{-i \phi} \\
      e^{+i \phi} & 0
    \end{array}\right)
  \right]
\end{equation}
to all qubits, obtaining the state $\rho(\phi)$. Then, the parity
\begin{equation}
P(\phi):=P_{\mathrm{even}}(\phi)-P_{\mathrm{odd}}(\phi),
\end{equation}
is obtained for various $\phi$ by repeated preparation and measurement in the computational basis. Here, $P_{\mathrm{even/odd}}(\phi)$ is the probability of finding an even/odd number of 1s in the output. The amplitude of $P(\phi)$ is obtained by fitting a sine to $P(\phi)$ (with free phase and amplitude).

We now show that, more precisely,
$\rho_{0\cdots 0,1\cdots 1}$ (without absolute values) is proportional to the highest frequency component of $P(\phi)$. What is more, the Fourier transform of $P(\phi)$ gives additional information about the anti-diagonal elements of $\rho$.

Given a state $\rho$, the parity $P=P_{\mathrm{even}}-P_{\mathrm{odd}}$ is given by the expectation value of $Z^{\otimes N}$. To see this, consider the case for two qubits,
\begin{align}
  Z\otimes Z&=(\ket{0}\!\bra{0}-\ket{1}\!\bra{1})\otimes(\ket{0}\!\bra{0}-\ket{1}\!\bra{1})\\
  &= \ket{0}\! \bra{0}\otimes \ket{0}\! \bra{0} - \ket{0}\! \bra{0} \otimes \ket{1}\! \bra{1}  - \ket{1}\! \bra{1} \otimes \ket{0}\! \bra{0} + \ket{1}\! \bra{1} \otimes \ket{0}\! \bra{0} \\
  & \rightarrow \ket{00}\! \bra{00} -  \ket{01}\!\bra{01} - \ket{10}\!\bra{10}  + \ket{11}\!\bra{11} \\
  & = \mathcal P_{\mathrm{even}} - \mathcal P_{\mathrm{odd}},
\end{align}
with $\mathcal P_{\mathrm{even/odd}}$ the projector onto the subspace with an even/odd number of 1s. Here, the arrow indicates a reordering of the tensor basis and an omission of the `$\otimes$', as is customary. With the above, it is straightforward to see that
\begin{align}
 Z^{\otimes N}=\mathcal P_{\mathrm{even}} - \mathcal P_{\mathrm{odd}}
\end{align}
in general, so that $\langle Z^{\otimes N}\rangle=P_{\mathrm{even}}-P_{\mathrm{odd}}$.

Now, note that
\begin{equation}
  U^\dagger(\phi) Z\,U(\phi)=
    e^{- i(\phi-\pi/2)}\ket{0}\!\bra{1}+
      e^{i(\phi-\pi/2)}\ket{1}\!\bra{0}
\end{equation}
so that
\begin{align}
  [U^\dagger(\phi) Z\, U(\phi)]^{\otimes N} & =\sum_l e^{i(\phi-\pi/2)(\lvert l \rvert-\lvert l \rvert_0)}\ket{l}\! \bra{\bar l},
\end{align}
with $\lvert l \rvert$ the number of 1s in $l$ (the Hamming weight of $l$) and $\lvert l \rvert_0=n-\lvert l \rvert $ the number of 0s in $l$. Thereby,
\begin{align}
P(\phi)= \langle Z^{\otimes N}\rangle_\phi&\equiv \mathrm{tr}\left[Z^{\otimes N} U_\phi^{\otimes N}\rho\   (U_\phi^\dagger)^{\otimes N}\right]\\
&=\sum_l e^{i (\phi-\pi/2)(\lvert l \rvert-\lvert l \rvert_0)} \rho_{\bar l,l}\\
&=\sum_{k'=-N}^{N} e^{i (\phi-\pi/2)k'} \rho^{(k')}\label{eq:pk}\\
&=m\, e^{- i \frac{2\pi}{m} N x} \left(\frac{1}{m}\sum_{k=0}^{m-1}e^{i  \frac{2\pi k}{m} x} \rho^{(k-N)}\right)\\
&=m\, e^{- i \frac{2\pi}{m}N x}\bar{\mathcal F}\left[\rho^{(\cdot-N)}\right](x),
\end{align}
with  $\rho^{(k')}$ the sum of all matrix elements of the form $\rho_{\bar l,l}$ such that $\lvert l \rvert-\lvert l \rvert_0=k'$, $k=k'+N$, $m=2N+1$, $x=\frac{(\phi-\pi/2)m}{2\pi}$,  and $\bar{\mathcal F}$ the inverse Fourier transform. Note that $P(\phi)$ is periodic with period $2\pi$.

By the shift theorem,
\begin{equation}
\mathcal F[P'](k)=m \rho^{(k)},
\end{equation}
with $P'(x)=P[2\pi(x/m+1/4)]$ and $\mathcal F[P'](k)=\sum_{x=0}^{m-1} e^{- i \frac{2\pi k}{m} x}P'(x)$ the Fourier transform of $P'$. 

In particular,
\begin{equation}
  \rho_{0\cdots 0,1\cdots 1}=\frac{\mathcal F[P'](N)}{2N+1}.
\end{equation}
Assuming that $\rho^{(k)}=0$ unless $k=\pm N$, which is generally not satisfied, we have by Eq. \eqref{eq:pk} that $P(\phi)=2\cos[(\phi-\pi/2)n+\theta]\lvert \rho_{0\cdots 0,1\cdots 1} \rvert$, with $\theta=\arg(\rho_{0\cdots 0,1\cdots 1})$, in which case the amplitude of $P(\phi)$ gives $2\lvert \rho_{0\cdots 0,1\cdots 1} \rvert$.

We now give an estimate of the sample complexity for obtaining $\rho_{0\cdots 0,1\cdots 1}$, ignoring experimental imperfections.
To resolve the highest frequency component of $P(\phi)$ requires $O(N)$ samples of $P(\phi)$. The variance of the estimator $E$ of the mean of a Gaussian random variable $R$, obtained by the average of $n$ samples from $R$ is $\sigma^2_{E}=\sigma_R^2/n$. Thus, estimating $P(\phi)$ to precision $\lvert \rho_{0\cdots 0,1\cdots 1}\rvert$ requires $n=\sigma_{P(\phi)}^2/\lvert\rho_{0\cdots 0,1\cdots 1}\rvert^2$ samples, with $\sigma_{P(\phi)}^2=\langle (Z^{\otimes N})^2 \rangle - \langle (Z^{\otimes N}) \rangle^2=O(1)$. This gives a total of $O(N/\lvert\rho_{0\cdots 0,1\cdots 1}\rvert^2)$ measurements to resolve $\rho_{0\cdots 0,1\cdots 1}$.

\vspace{0.5cm}
{\Large\textbf{IV.\quad Multiple quantum coherences}}
\vspace{0.5cm}

In the technique of multiple quantum coherences (MQC), the signal $S(\phi)$, analogous to $P(\phi)$ in the case of parity oscillations, is defined as
\begin{equation}\label{eq:Sphi}
S(\phi)=\mathrm{tr}[\rho\rho(\phi)].
\end{equation}
Here, $\rho(\phi)$ is obtained from $\rho\equiv \rho(0)$ by applying a local RZ-rotation, with angle $\phi$, to each qubit, 
\begin{equation}
\rho(\phi)=e^{-i\phi \sum_{i}Z_i/2}\rho\, e^{i\phi \sum_{i}Z_i/2}.
\end{equation}

Similar to parity oscillations, $S(\phi)$ contains information about sums of the elements of $\rho$, which is revealed by taking the Fourier transform of $S(\phi)$. This has been rederived many times, e.g. in Ref. \cite{wei2020verifying}, but we repeat the derivation here for a clearer comparison with the analogous derivation for parity oscillations and in a notation that we consider unambiguous. 

Consider the projector $P_m$ onto the subspace containing those states that have eigenvalue $m$ of the total spin operator $Z^{\mathrm{tot}}=\sum_i Z_i$. This space is spanned by all computational basis states $\ket x$ for which the difference between the number of 0s and the number of 1s is $m$. First, define
\begin{equation}
    \rho_q=\sum_m P_m \rho P_{m-q}.
\end{equation}
It can be straightforwardly verified that
\begin{equation}
    \mathrm{tr}(\rho_q\rho_p)=\delta_{q,-q}\mathrm{tr}(\rho_{q}\rho_{-q})
\end{equation}
and 
\begin{equation}
e^{-i\phi \sum_{i}Z_i/2}\rho_q\, e^{i\phi \sum_{i}Z_i/2}=e^{-i\phi q/2}\rho_q.
\end{equation}

Since $\sum_m P_m=\mathbb 1$, we may expand $\rho$ as  $\rho=\sum_{m,m'}, P_m \rho P_{m'}=\sum_q \rho_q$. Thus,
\begin{align}
    S(\phi)&=\sum_{q,p}\mathrm{tr}(e^{-i\phi q/2}\rho_q \rho_{p})\\
    &=\sum_{q=-N}^{N} e^{-i\phi q/2}\mathrm{tr}(\rho_q\rho_{-q}).
\end{align}
Thereby, 
\begin{equation}
I_q\equiv\mathrm{tr}(\rho_q\rho_{-q})=\frac{\mathcal F[S'](q)}{2N+1},
\end{equation}
with $S'(x)=S\left(-\frac{4\pi x}{2N+1}\right)$. In particular, 
\begin{equation}
    I_N=I_{-N}=\lvert \rho_{0\cdots 0,1\cdots1}\rvert^2
\end{equation}
so that (the absolute value squared) of the far off-diagonal matrix element of $\rho$ can be estimated by the amplitude of the highest frequency component of the signal $S(\phi)$. 

Since $\rho$ is Hermitian, $S(\phi)$ can be estimated by measuring the `observable' $\rho$ in the state $\rho(\phi)$, but this requires a noiseless circuit mapping the unknown eigenbasis of $\rho(\phi)$ to the computational basis. Alternatively, there exists the following protocol. (1) Starting from the initial state $r_0=\ketbra{0\cdots 0}{0\cdots 0}$, prepare $\rho$ using the noisy circuit $\tilde U$. (2) Apply an RZ-rotation with angle $\phi$ to each qubit. (3) Undo the state preparation circuit. (4) Measure the probability to return to the state $\ket{0\cdots 0}$. 

Assuming perfect gates and measurement, the above protocol estimates
\begin{align}
\mathrm{tr}[U^\dagger\,RZ^{\otimes N}(\phi)\,U\,r_0\,U^\dagger\,RZ^{\otimes N}(-\phi)\,U\,r_0]&=\mathrm{tr}[RZ^{\otimes N}(\phi)\,U\,r_0\,U^\dagger\,RZ^{\otimes N}(-\phi)\,U\,r_0\,U^\dagger]\\
&=\mathrm{tr}[\rho(\phi)\rho]\\
&=S(\phi). 
\end{align}

In Ref.~\cite{gaerttner2018relating}, it was shown that the protocol remains valid when noise can be modeled as a Markovian bath with local absorption and emission at equal rates, along with local dephasing, and no idling time. Although this assumption may be justified in NMR experiments, it generally does not capture noise processes in quantum computing hardware and presupposes uncorrelated noise.

\end{document}